\documentclass[11pt,a4paper]{article}
\pdfoutput=1
\usepackage{jheppub}
\usepackage[utf8]{inputenc}
\usepackage[english]{babel}
\usepackage[T1]{fontenc}
\usepackage{hyperref}
\usepackage{amsmath}
\usepackage{amsfonts}
\usepackage{dsfont} 
\usepackage{color} 
\usepackage{caption}
\usepackage{bbold}

\newcommand{\Tr}{\mathrm{Tr}}

\usepackage{enumerate}
\usepackage{subfig}
\usepackage{graphicx}

\usepackage{csquotes}

\definecolor{mygray}{gray}{0.3}

\newcommand\beq{\begin{equation}}
\newcommand\eeq{\end{equation}}
\newcommand{\bes}{\begin{eqnarray}}
\newcommand{\ees}{\end{eqnarray}}

\newcommand\restr[2]{{
  \left.\kern-\nulldelimiterspace 
  #1 
  \vphantom{\big|} 
  \right|_{#2} 
  }}

\usepackage{multicol}
\usepackage{amssymb}
\usepackage{graphicx}

\usepackage[normalem]{ulem}

\newcommand{\be}{\begin{equation}}
\newcommand{\ee}{\end{equation}}
\newcommand{\bea}{\begin{eqnarray}}
\newcommand{\eea}{\end{eqnarray}}

\begin{document}
\author[a]{Astrid Eichhorn,} 
\author[b]{Antonio D. Pereira,} 
\author[c,d]{Andreas G. A. Pithis} 

\affiliation[a]{CP3-Origins, University of Southern Denmark, Campusvej
  55, DK-5230 Odense M, Denmark} 
    \affiliation[b]{Instituto  de  F\'{i}sica,  Universidade  Federal  Fluminense,\\ Campus  da  Praia  Vermelha,  Av.   Litor\^{a}nea  s/n,  24210-346,  Niter\'{o}i,  RJ,  Brazil}
 \affiliation[c]{Institut f\"{u}r
  Theoretische Physik, Universit\"{a}t Heidelberg, Philosophenweg 16,
  69120 Heidelberg, Germany}
  \affiliation[d]{Scuola  Internazionale  Superiore  di Studi Avanzati (SISSA),\\ via, Bonomea, 265, 34136 Trieste, Italy}

\emailAdd{eichhorn@cp3.sdu.dk}
\emailAdd{adpjunior@id.uff.br}
\emailAdd{pithis@thphys.uni-heidelberg.de}
 
\title{The phase diagram of the multi-matrix model with $ABAB$ interaction from functional renormalization
}

\begin{abstract}
{At criticality, discrete quantum-gravity models are expected to give rise to continuum spacetime.
Recent progress has established the functional renormalization group method in the context of such models as a practical tool to study their critical properties and to chart their phase diagrams. Here, we apply these techniques to the multi-matrix model with $ABAB$ interaction potentially relevant for Lorentzian quantum gravity in 3 dimensions. We characterize the fixed-point structure and phase diagram of this model, paving the way for functional RG studies of more general multi-matrix or tensor models encoding causality and subjecting the technique to another strong test of its performance in discrete quantum gravity by comparing to known results.}
\end{abstract}

\maketitle

\section{Introduction}
Understanding the quantum properties of spacetime is a fascinating challenge. A variety of approaches is currently being explored concertedly; more recently with an increased interest in understanding relations between different perspectives. The matrix/tensor model approach~\cite{DiFrancesco:1993cyw,Gurau:2016cjo,gurau2017random,Rivasseau:2016wvy,Delporte:2018iyf} is located at a particular vantage point within this \enquote{landscape} of theories, with potential links to a number of different ones: Firstly, its origin in two-dimensional gravity is closely linked to string theory. 
Secondly, its generalization to higher dimensions is closely connected to a model that is being explored in the context of the AdS/CFT conjecture~\cite{Witten:2016iux,Rosenhaus:2019mfr}. Thirdly, in matrix models a tentative connection to an asymptotically safe fixed point in the vicinity of two dimensions has been found~\cite{Eichhorn:2014xaa} and conjectured in higher dimensions~\cite{Eichhorn:2019hsa}. Fourth, this class of models provides a combinatorial approach to dynamical triangulations, complementing computer simulations of the latter~\cite{Ambjorn:2001cv,Ambjorn:2012jv,Loll:2019rdj}. This rich set of connections motivates a joint study of matrix and tensor models and in particular the search for a universal continuum limit in these. Functional Renormalization Group techniques have been developed for these models~\cite{Eichhorn:2013isa,Eichhorn:2014xaa,Eichhorn:2017xhy} and are being applied in~\cite{Eichhorn:2018ylk,Eichhorn:2018phj,Lahoche:2019ocf,Eichhorn:2019hsa} and related tensorial group field theories~\cite{Benedetti:2014qsa,Benedetti:2015yaa,Geloun:2015qfa,Geloun:2016qyb,Carrozza:2016tih,Lahoche:2016xiq,Lahoche:2018vun,Lahoche:2018oeo,BenGeloun:2018ekd,Lahoche:2019orv,Baloitcha:2020idd,Pithis:2020sxm,Pithis:2020kio}, with the potential to discover a universal continuum limit beyond perturbation theory, see also~\cite{Krajewski:2015clk,Krajewski:2016svb} for related developments with the Polchinski equation.\\
To set the stage for our studies, we first provide a more in-depth overview of the relevant quantum-gravity approaches.

The above approaches center on the path integral
\begin{equation}\label{LorentzianPI}
    Z=\int \mathcal{D}g~\text{e}^{i S[g]}\,,
\end{equation}
where the integration runs over all field histories, given by spacetime metrics $g$ of the $d$-dimensional manifold $\mathcal{M}$ up to diffeomorphisms thereof at fixed spacetime topology. 
Working with the Einstein-Hilbert action, one can make sense of the weak-field limit in an effective-field theory framework~\cite{Donoghue:2012zc}, but encounters perturbative non-renormalizability, resulting in a loss of predictivity, beyond~\cite{tHooft:1974toh,Goroff:1985th}. 
A possible way out of this issue is to replace the Einstein-Hilbert action by one which allows for a unitary and perturbatively renormalizable QFT. However, this might dispense with micro-causality as in the case of higher-derivative gravity~\cite{PhysRevD.16.953,Donoghue:2019ecz,Donoghue:2019fcb} or with Lorentz invariance as in Ho\v{r}ava-Lifshitz gravity~\cite{Horava:2009uw}, see~\cite{Steinwachs:2020jkj} for a recent review. 

An alternative pathway to quantize gravity within the continuum formulation of the path integral is explored by the asymptotic-safety program. To sidestep the problems of the perturbative quantization, this approach is based on an interacting fixed point in the Renormalization Group (RG) flow for gravity in the UV~\cite{weinberg1978critical,weinberg1979ultraviolet,Reuter:1996cp}. If it exists, such a fixed point provides a well-defined continuum limit to the path integral. At the same time, it generalizes perturbative renormalizability by ensuring that the low-energy limit is parameterized by only a finite set of free parameters, namely  the relevant directions of the fixed point. There are several techniques suitable to explore asymptotic safety in gravity, falling into the two broad categories of lattice approaches and continuum approaches.
A much used method pioneered for gravity by Reuter~\cite{Reuter:1996cp} is provided by the functional Renormalization Group (FRG)~\cite{Wetterich:1992yh,Morris:1993qb}, see~\cite{Dupuis:2020fhh} for a review. At its core lies the implementation of the Wilsonian idea of a coarse graining operation which progressively eliminates short scale fluctuations. Indeed, all explicit calculations within truncated RG flows find evidence for the existence of such a fixed point, defining the Reuter universality class, providing compelling indications for asymptotic safety in Euclidean gravity, see, e.g.,~\cite{Percacci:2017fkn,Eichhorn:2018yfc,reuter2018quantum,Pereira:2019dbn,Reichert:2020mja,Bonanno:2020bil,Eichhorn:2020mte} for recent reviews and introductions. Rephrased in a lattice-like language, such a universality class enables one to take a universal continuum limit.\\
Open questions in this approach have been discussed in~\cite{Donoghue:2019clr,Bonanno:2020bil} and include the fate of background independence, given the assumption of an auxiliary background metric therein~\cite{Reuter:1996cp,Becker:2014qya,Falls:2020tmj}. Moreover, since the signature of the setup is Euclidean and one can in general not expect the Wick rotation to exist in a quantum gravitational context~\cite{Demmel:2015zfa,Visser:2017atf,Baldazzi:2018mtl,Baldazzi:2019kim}, it is open how to relate these results and in particular the feature of asymptotic safety to Lorentzian quantum gravity; for a first step in this direction see~\cite{Manrique:2011jc}. Similarly to the characterization of other interacting fixed points, a concerted use of several different techniques could in the future provide a qualitatively and quantitatively robust grasp of the fixed point and its properties. In the case of quantum gravity, background-independent, Lorentzian approaches to quantum gravity are of particular interest to explore as techniques that can complement the FRG results.

One such promising approach to evaluate the path integral over geometries, possibly extended by a sum over topologies, and to discover a universal continuum limit, a.k.a.~asymptotic safety, is by means of a sum over discrete triangulations, together with exchanging the continuum action with its discretized reformulation. In spite of the physical discreteness that some of these settings exhibit, such as, e.g., Loop Quantum Gravity~\cite{Rovelli:1987df,Rovelli:1989za,Thiemann:2007pyv}, taking a universal continuum limit is a central goal also in this setting. For instance, in group field theory~\cite{Freidel:2005qe,Oriti:2006se,Carrozza:2016vsq,Pithis:2019tvp}, promising results regarding asymptotic freedom~\cite{BenGeloun:2012pu,BenGeloun:2012yk} and asymptotic safety~\cite{Carrozza:2014rya,Carrozza:2016tih} have been obtained, see~\cite{Carrozza:2016vsq} for a review. Similarly, RG techniques are being developed and applied in the search for a critical point in spin foams, see~\cite{Dittrich:2014ala, Steinhaus:2020lgb}. In this setting, studies of the phase diagram of quantum gravity have only recently started~\cite{Delcamp:2016dqo,Bahr:2016hwc,Steinhaus:2018aav,Bahr:2018gwf}. In contrast, in the Euclidean and Causal Dynamical Triangulation approaches (EDT, CDT)~\cite{Loll:1998aj,Ambjorn:2012jv,Ambjorn:2013tki,Loll:2019rdj}, much is already known about the phase diagram. Going beyond two dimensions, early numerical studies using Monte-Carlo methods~\cite{Ambjorn:1991pq,Agishtein:1991cv,Catterall:1994pg,Bilke:1996ek} have only recovered unphysical geometries~\cite{Bialas:1996wu,deBakker:1996zx,Catterall:1996gj} in the Euclidean setting, though the inclusion of additional terms in the action or measure of the path integral, associated with additional tunable parameters, might change the situation~\cite{Laiho:2016nlp,Ambjorn:2013eha}. It is argued that these pathological configurations are the result of topology change leading to spaces called branched polymers which are built from one-dimensional branched-out filaments~\cite{Ambjorn:1993sy,Ambjorn:1995dj,Bialas:1996wu,Catterall:1994pg}. At the classical level, as long as Morse geometries are excluded~\cite{Sorkin:1997gi,Dowker:1997hj,Borde:1999md}, it is long known that topology change leads to a degenerate local light cone structure and thus to a violation of micro-causality~\cite{Geroch:1967fs}. 
Thus, in the CDT approach discrete spacetime configurations with spatial topology change are excluded~\cite{Ambjorn:1998xu}. This leads to a much better behaved theory with the potential to produce a phase with physically relevant, i.e., extended macroscopic geometries~\cite{Ambjorn:2012jv,Ambjorn:2013tki,Loll:2019rdj}, bordered by a second-order phase transition~\cite{Ambjorn:2011cg,Ambjorn:2012ij,Ambjorn:2016mnn,Ambjorn:2020azq} that enables a continuum limit. Yet, a key challenge in the numerical approach to dynamical triangulations remains to follow RG trajectories towards the continuum limit~\cite{Ambjorn:2014gsa} and calculate the scaling spectra to determine and characterize the universality class.

Matrix~\cite{DiFrancesco:1993cyw} and tensor models~\cite{Gurau:2016cjo,gurau2017random,Rivasseau:2016wvy,Delporte:2018iyf} sit at a confluence of several of these approaches. They encode random discrete geometries in dimensions $d\geq 2$. Their general idea is to represent  {$d-1$ simplices corresponding to building blocks of geometry as} rank $d$-tensors. The tensor action encodes how to glue these building blocks together to construct $d$-dimensional discrete geometries. These correspond to the Feynman diagrams in the perturbative expansion of the tensor path integral. This establishes a duality between tensor models and the discrete gravitational path integral. Indeed, in their simplest form matrix and tensor models can be understood as generators of EDTs. However, apart from the case in $d=2$~\cite{DiFrancesco:1993cyw}, their continuum limit so far only leads to geometrically degenerate configurations~\cite{Gurau:2016cjo,gurau2017random}, the same way as in EDT, or planar ones~\cite{Lionni:2017xvn}. Recent results indicate the possibility for non-trivial universality classes~\cite{Eichhorn:2018ylk,Eichhorn:2019hsa}; however, the geometric properties of the corresponding phases have not been investigated yet. If a universal continuum limit within a phase with desirable geometric properties can be taken, asymptotic safety can be confirmed in a background-independent fashion and with straightforward access to the scaling exponents. Thus, the critical role of universality in these models has been emphasized in~\cite{Rivasseau:2011hm,Eichhorn:2018phj}.\\
Yet, in studies beyond two dimensions, the inclusion of causality has remained an open question -- similarly as in the continuum approach to asymptotic safety.
The success of CDT may be taken as a hint to consider additional structure enforcing micro-causality to recover higher-dimensional physical continuum geometries from such models. This idea has already been implemented in the context of matrix models for $2d$ quantum gravity, giving rise to a description equivalent to CDT in $(1+1)$ dimensions~\cite{Benedetti:2008hc}. In order to explore the impact of Lorentzian signature in higher dimensions, one could naturally try to impose such causality conditions on a model for tensors of rank $d\geq 3$. Yet, there already exists a proposed correspondence between a multi-matrix model with CDT in $2+1$ dimensions~\cite{Ambjorn:2001br}.  More precisely, it corresponds to a Hermitian two-matrix model with $ABAB$ interaction which also has intimate connections to vertex-models of statistical physics~\cite{ZinnJustin:2003kq,Kazakov:1998qw,ZinnJustin:1999wt,Kostov:1999qx}. It leads to a variant of CDT defined on an enlarged ensemble of configurations which also allows for specific degeneracies of the local geometry, dubbed \enquote{wormholes} in the literature~\cite{Ambjorn:2001br}. The purpose of this article is to chart the phase structure of this matrix model and in particular to study its continuum limits by means of the FRG methodology. 
The application of the FRG in the discrete quantum gravity context facilitates a background-independent form of coarse graining where the number of degrees of freedom serves as a scale for a Renormalization Group flow. This program was started by analyzing matrix models for $2d$ Euclidean quantum gravity in~\cite{Eichhorn:2013isa, Eichhorn:2014xaa} and has by now been extended to tensor models for Euclidean quantum gravity in $3$ and $4$ dimensions~\cite{Geloun:2016xep,Eichhorn:2017xhy,Eichhorn:2018ylk,Eichhorn:2019hsa}, see~\cite{Eichhorn:2018phj} for a review. Related developments for non-commutative geometry~\cite{Sfondrini:2010zm,perezsanchez2020multimatrix} and tensorial group field theories~\cite{Benedetti:2014qsa,Benedetti:2015yaa,Geloun:2015qfa,Geloun:2016qyb,Carrozza:2016tih,Lahoche:2016xiq,Lahoche:2018vun,Lahoche:2018oeo,BenGeloun:2018ekd,Lahoche:2019orv,Baloitcha:2020idd,Pithis:2020sxm,Pithis:2020kio} exist.
Recently, a first FRG analysis of the above-mentioned causal matrix model for CDT in $1+1$ dimensions has been completed~\cite{CMMFRG}. 

The article is organized as follows: In Sec.~\ref{sec:causality} we discuss causality in the matrix-model context and review the relation of the $ABAB$ matrix model to CDTs, following Refs.~\cite{Ambjorn:2001br,Ambjorn:2003sr,Ambjorn:2003ct}. In Sec.~\ref{sec:FRGABAB}, we briefly review functional RG techniques and apply them to the $ABAB$ matrix model. We then present results for the phase diagram and fixed-point structure. In Sec.~\ref{sec:reviewpd} we review what is known about the phase diagram in the literature~\cite{Ambjorn:2001br,Ambjorn:2003sr,Ambjorn:2003ct,Kazakov:1998qw,ZinnJustin:2003kq} and compare our results to it. Finally, in Sec.~\ref{sec:discussion} we discuss implications of our results and future directions.

\section{Causality and matrix models}\label{sec:causality}
Spacetime is rather distinct from space, both at the conceptual as well as mathematical level. Therefore, it is crucial to take this difference into account in quantum gravity, with the distinct phase diagrams of CDT and EDT constituting a clear example of the impact of causality. In the matrix and tensor model approach, the additional structure imposed on discrete configurations by causality can be implemented through additional degrees of freedom: In Refs.~\cite{Benedetti:2008hc,CMMFRG}, this is done by an external matrix, whereas the $ABAB$ model uses two dynamical matrices to generate configurations which carry imprints of causality. This strongly motivates the further development of FRG techniques for multi-matrix/tensor settings, such that similar developments can be made possible in higher dimensions.\\
In this section, we will review the relation of the $ABAB$ matrix model to 2+1 dimensional discrete spacetime configurations. In particular, we will follow Refs.~\cite{Ambjorn:2001br,Ambjorn:2003sr,Ambjorn:2003ct} to also review the connection to CDT.

\begin{figure}[!t]
\includegraphics[angle=0,width=13cm]{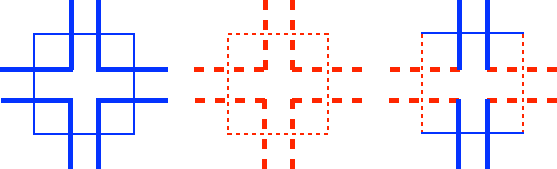}
\centering
\caption{Distinct squares correspond to blue/continuous (type $A^4$), red/dashed (type $B^4$) and bi-colored/continuous-dashed (type $ABAB$) $4$-vertices in the dual ribbon/fat graph generated by the matrix model.}\label{figure:qvcorrespondence}
\end{figure}

The $ABAB$ matrix model is defined by the following partition function
\begin{eqnarray}\label{mmmparititionfunction}
Z[J_A, J_B] &=&\int \text{d}A\, \text{d}B\, {\rm exp}\Big(
 -\frac{1}{2}\Tr\left(AA\right)-\frac{1}{2}\Tr\left(BB\right)+\frac{\bar{\alpha}_1}{4}\Tr\left(AAAA\right)+\frac{\bar{\alpha}_2}{4}\Tr\left(BBBB\right)\nonumber\\
 &{}&+\frac{\bar{\beta}}{2}\Tr\left(ABAB\right)  + \Tr \left(J_A A\right) + \Tr \left(J_B B\right)\Big),
\end{eqnarray}
where $A$ and $B$ are Hermitian $N'\times N'$ matrices and $J_A$ and $J_B$ are the respective external ($N^\prime \times N^\prime$-matrix) sources. Its Feynman diagrams are ribbon diagrams with two distinct types of lines; such that the duals of the three types of vertices correspond to three distinct squares, cf.~Fig.~\ref{figure:qvcorrespondence}. This already highlights that the model reduces to the standard two-dimensional gravity case, when $\bar{\beta} \rightarrow 0$. In the presence of $\bar{\beta}$, 2+1 dimensional structure can be encoded~\cite{Ambjorn:2001br}.

\begin{figure}[ht!]
\includegraphics[angle=0,width=13cm]{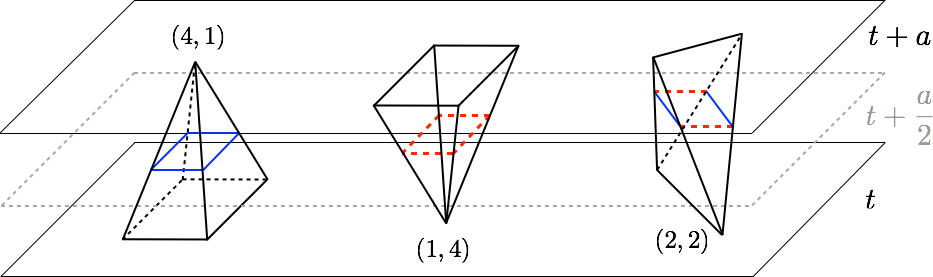}
\centering
\caption{The three fundamental building blocks of CDT in $2+1$ dimensions which interpolate between two consecutive spatial hypersurfaces at integer times $t$ and $t+a$, cf.~Fig.~1 in Ref.~\cite{Ambjorn:2001br}. The numbers at the pyramids and the tetrahedron refer to the number of vertices in the quadrangulations at constant integer time $t$ and $t+a$. In between, their intersections with the $t+\frac{a}{2}$-plane are shown, giving rise to an equilateral quadrangulation thereof in terms of blue, red and bi-colored squares. The colorization emphasizes that $2+1$ dimensional information is encoded in a $2d$ setting \textit{with colors}.}\label{figure:CDT21}
\end{figure}

\begin{figure}[ht!]
\includegraphics[angle=0,width=8.5cm]{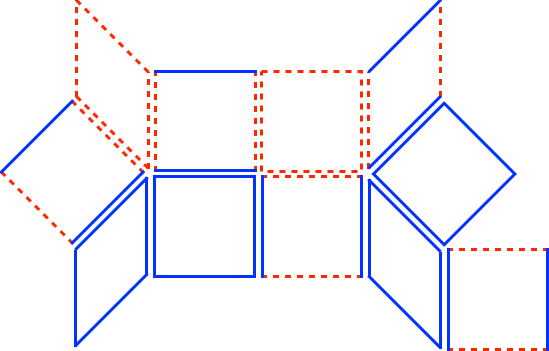}
\centering
\caption{Snapshot of a quadrangulation of the intermediate $t+\frac{a}{2}$-plane, cf.~Fig.~2 in Ref.~\cite{Ambjorn:2001br}.}
\label{figure:quadrangulation}
\end{figure}

In CDTs, one only considers geometries with Lorentzian signature which admit a global foliation in proper time $t$ and disallow topology change such that micro-causality is rigidly maintained. The basic building blocks are pyramids and tetrahedra with distinct timelike and spacelike edges, of which three distinct types make up the discrete configuration, cf.~Fig.~\ref{figure:CDT21}, see also Refs.~\cite{Ambjorn:2000dja,Ambjorn:2001br,Kazakov:1989bc,Ambjorn:2000dv}. The correspondence with the ribbon graphs of the $ABAB$ matrix model arises when considering the spatial hypersurface at $t+a/2$, also indicated in Fig.~\ref{figure:CDT21}, where the three distinct types of squares dual to the vertices of the matrix model, cf.~Fig.~\ref{figure:qvcorrespondence}, appear. In the $t+a/2$ spatial planes, quadrangulations are formed as, e.g., in Fig.~\ref{figure:quadrangulation}, the duals of which are the ribbon graphs of the matrix model. This already suggests that any CDT configuration can be encoded in terms of a Feynman diagram of the matrix model, which would motivate setting the CDT partition function equal to the free energy of the matrix model, as usual in the correspondence between triangulations and matrix models.  To see this in more detail and further discuss whether or not there is an exact correspondence, let us follow the discussion in~\cite{Ambjorn:2001br}.

Indeed, the entire information on the CDT partition function can be encoded in the one-step propagator which provides the transition amplitude between the spatial hypersurfaces at $t$ and $t+a$. 
It is this property that enables a connection to the $ABAB$ matrix model. Indeed, it has been shown in~\cite{Ambjorn:2001br} that the Euclideanized transition amplitude between the triangulation $\Delta$ of the spatial hypersurface $\Sigma_t$ and  $\Delta(\Sigma_{t+a})$ is given by (cf.~Eq.~6 in~\cite{Ambjorn:2001br})
\begin{align}
\langle \Delta(\Sigma_{t+a})\vert \hat{T} \vert \Delta(\Sigma_t)\rangle& =\langle N_{14}\vert\hat{T}\vert N_{41}\rangle\nonumber\\&=\text{e}^{-(\lambda-\kappa)(N_{41}+N_{14})}\sum_{N_{22}=N_{22}^{\text{min}}}^{N_{22}^{\text{max}}}\mathcal{N}_{\text{CDT}}(N_{41},N_{14},N_{22})\text{e}^{-(\frac{1}{2}\lambda+\kappa)N_{22}},
\end{align}
where $\hat{T}$ is the transfer matrix. In the above expression, $N_{41}$ is the number of squares at $t$ and there are $N_{14}$ squares in $t+a$. The sum is over all intermediate quadrangulations, and $\mathcal{N}_{\text{CDT}}(N_{41},N_{14},N_{22})$ denotes the total number of quadrangulations at fixed $N_{22}$. We note that this expression holds for spaces of spherical topology triangulated by a large number of squares. Further, the above expression assumes the discretized Einstein-Hilbert action, such that $\lambda$ and $\kappa$ are related to the bare cosmological constant $\Lambda$ and bare Newton coupling $G_N$ as follows
\begin{equation}
    \kappa = \frac{a}{4\pi G_{\text{N}}}\left(-\pi+3\arccos{\frac{1}{3}}\right),~~~\lambda=\frac{a^3 \Lambda}{24\sqrt{2}\pi},
\end{equation}
and $a$ is the lattice spacing.

Finally, introducing the dimensionless boundary constant $z_t$ understood as a cosmological term for the boundary area, one can set up the discrete Laplace transform of the transfer matrix,
\begin{equation}
    \langle z_{t+a}\vert \hat{T}\vert z_t\rangle = \sum_{N_{41},N_{14}}\text{e}^{-z_t N_{41}-z_{t+a}N_{14}}\langle N_{14}\vert\hat{T}\vert N_{41}\rangle,
\end{equation}
which together with the identifications
\begin{equation}\label{eq:ccrelations}
    \bar{\alpha}_1=\text{e}^{\kappa-\lambda-z_t},~~~ \bar{\alpha}_2=\text{e}^{\kappa-\lambda-z_{t+a}},~~~\bar{\beta}=\text{e}^{-\frac{1}{2}\lambda+\kappa},
\end{equation}
makes the close relation of the free energy of the matrix model,
\be
N'^2 F=- \ln Z[0,0]=\sum_{N_{41}, N_{14}, N_{22}}\mathcal{N}_{\text{MM}}(N_{41},N_{14}, N_{22})\bar{\alpha}_1^{N_{41}} \bar{\alpha}_2^{N_{14}} \bar{\beta}^{N_{22}},
\ee
to the CDT-partition function obvious. 

\begin{figure}[ht!]
\includegraphics[angle=0,width=7.5cm]{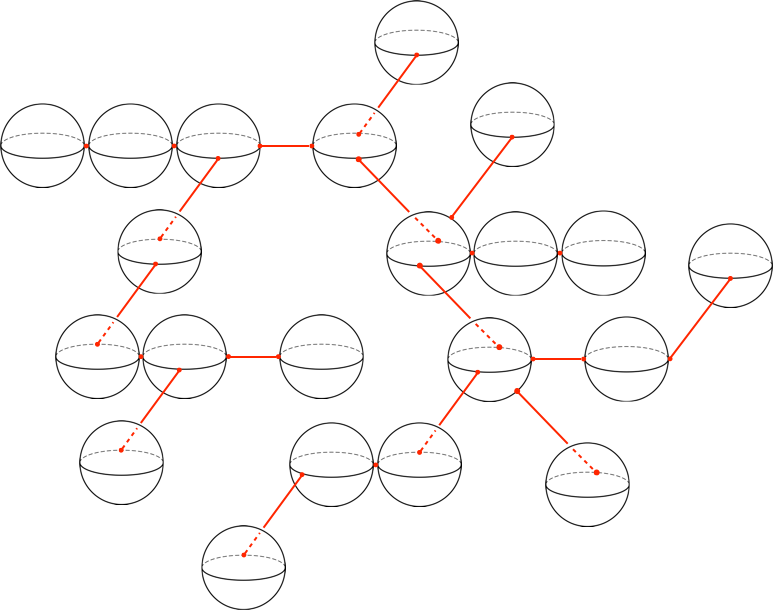}
\centering
\caption{Snapshot of a network of spheres at time $t$ which touch each other at most once at points or are connected by single edges of the quadrangulation caused by so-called touching interactions in the dual matrix model description at $t+\frac{a}{2}$.}\label{figure:baby}
\end{figure}

However, the number of configurations generated by the CDT model is smaller than that generated by the matrix model, i.e., $\mathcal{N}_{\text{MM}}>\mathcal{N}_{\text{CDT}}$: The matrix model generically generates disconnected subgraphs via so-called touching interactions (see below). This is most easily seen by switching off the interactions in one sector, $\bar{\alpha}_2\rightarrow 0$ and integrating out $B$. The resulting single-matrix model contains multi-trace terms.
In the dual triangulation picture, these lead to branched trees of spherical bubbles, see Fig.~\ref{figure:baby}. Clearly, in the generalized situation where $\bar{\alpha}_2$ is reinstated, touching interactions will also be present and yield such undesirable quadrangulations. Such pathologies of the local geometry are disallowed by construction in CDT~\cite{Ambjorn:2012jv}.

Despite these differences, the similarities between the $ABAB$ matrix model and CDTs reinforce the more general notion that causality can be imposed on matrix and tensor models by enlarging the field content of the model and introducing a second matrix/tensor that ultimately enables a distinction of spacelike and timelike edges in the dual triangulation. This motivates us to perform a functional RG analysis of the $ABAB$ matrix model. On the one hand, a comparison to existing results in the literature will enable us to conduct a novel, powerful test of the performance of this technique. On the other hand, this will pave the way for future studies of multi-field models that encode causality in the interaction structure.

\section{FRG analysis of the $2$-matrix model with $ABAB$ interaction}\label{section:FRGanalysis}\label{sec:FRGABAB}

\subsection{FRG method for matrix models}\label{sec:FRGmethodformatrices}

The FRG is a powerful and versatile tool to implement the Wilsonian renormalization program. In  a standard, local field-theoretic setting, given the Euclidean path integral, one introduces a regulator function which suppresses the functional integration of modes below a given momentum cutoff $k$ which correspond to the slow modes in the Wilsonian perspective. By progressively lowering the values of $k$, one carries out the complete integration over all modes. Hence, instead of performing the path integral at once, it is computed in a momentum-shell-wise fashion. The central object in the FRG is the so-called flowing action $\Gamma_k$ which interpolates between the classical action $S$ (when $k\to\infty$) and the full effective action $\Gamma$ (when $k\to 0$). It satisfies a flow equation~\cite{Wetterich:1992yh,Morris:1993qb} which has a simple one-loop structure, making it very efficient for practical calculations, where approximations have to be employed. As the full propagator enters, nonperturbative physics is captured, despite the one-loop structure. At the formal level, the equation is exact and no approximation enters its derivation, rendering it formally equivalent to the path integral. For a recent review of the FRG in a broad range of contexts from condensed matter to quantum gravity, see~\cite{Dupuis:2020fhh}.

The Wilsonian picture as described above relies on a background structure which provides a notion of  momentum scales. Such a momentum scale is used as a coarse-graining parameter. In the case of matrix models for quantum gravity, there is no such background. Indeed, these models can be thought of as pre-geometric, with a smooth spacetime and notions of distance emerging in the continuum limit.
Yet, as introduced in Ref.~\cite{Brezin:1992yc}, a notion of renormalization group can naturally be defined if the number of entries of the matrices is taken as the coarse-graining parameter, see also Ref.~\cite{Bonnet:1998ei} for an extension of this idea to the case of two-matrix models. Hence, integrating out ``fast modes" corresponds to integrating out the outermost rows and columns. In 
Ref.~\cite{Eichhorn:2013isa}, this idea was translated into an exact flow equation for matrix models, paving the way for similar developments in group field theories~\cite{Benedetti:2014qsa} and tensor models~\cite{Eichhorn:2017xhy}. More recent developments can be found in Refs.~\cite{Eichhorn:2018ylk,Eichhorn:2019hsa,Lahoche:2019ocf,perezsanchez2020multimatrix}, see also~\cite{Eichhorn:2018phj} for a review. For a matrix model defined by the partition function,
\begin{equation}
Z[J] = \int \mathrm{d}M\,\mathrm{e}^{-S[M]+\Tr\left(J M\right)}\,,
\label{mmfe1}
\end{equation}
where $M$ denotes a Hermitian $N^\prime \times N^\prime$ matrix, $J$ an external ($N^\prime \times N^\prime$-matrix) source and $S[M]$ the classical action of the model, one adds a regulator function $\Delta S_N$ to the Boltzmann factor of the partition function and obtains the scale-dependent partition function $Z_N [J]$,
\begin{equation}
Z_N [J] = \int \mathrm{d}M\,\mathrm{e}^{-S[M]-\Delta S_N+\Tr\left(J M\right)}\,.
\label{mmfe2}
\end{equation}
We demand that the regulator $\Delta S_N$ has the structure of
\begin{equation}
\Delta S_N = \frac{1}{2}\sum_{a,b,c,d} M_{ab}R^{ab,cd}_N(a,b) M_{cd}\,,
\label{mmfe3}
\end{equation}
with $R_N$ being independent of the random matrix $M$. It is required to satisfy the following three properties:
\begin{align}
&\lim_{a/N\to 0,~b/N\to 0} R_{N}(a,b)_{ab,cd}>0,\label{Rproperties1}\\
&\lim_{N/a\to 0,~N/b\to 0}R_{N}(a,b)_{ab,cd}=0,\label{Rproperties2}\\
&\lim_{N\to N'\to\infty}R_{N}(a,b)_{ab,cd}\label{Rproperties3}\to\infty,
\end{align}
which suppress the matrix entries in the block $a,b=1,...,N$ and facilitate that the \enquote{UV} matrix entries with $a,b>N$ are integrated out~\cite{Eichhorn:2013isa}. It is then a formal manipulation to show that the flowing action\footnote{We adopt a \enquote{sloppy} notation where the argument of $\Gamma_N$ is denoted by $M$ but actually represents $\langle M \rangle_J$.} $\Gamma_N [M]$ defined by 
\begin{equation}
\Gamma_N [M] = \mathrm{sup}_{J}\left(\Tr(J M) - \mathrm{ln}~Z_N\right)-\Delta S_N\,,
\label{mmfe4}
\end{equation}
satisfies
\begin{equation}
\partial_t \Gamma_N = \frac{1}{2}\mathrm{Tr}\left[\left(\frac{\delta^2\Gamma_N}{\delta M \delta M}+R_N\right)^{-1}\partial_t R_N\right]\,.
\label{mmfe5}
\end{equation}
Indices were suppressed for simplicity. The $\mathrm{Tr}$ represents a sum over the indices and $\partial_t\equiv N\partial_N$. Such a derivative should actually be a finite difference at finite $N$. However, we are interested in the large-$N$ limit, justifying the explicit use of the derivative. The flowing action $\Gamma_N$ contains all terms compatible with the symmetries of the model. Thus, it can be expanded as
\begin{equation}
\Gamma_N [M] = \sum_I \bar{g}_I\,\mathcal{O}^I [M]\,,
\label{mmfe6}
\end{equation}
with $\mathcal{O}[M]$ denoting all the operators satisfying a given symmetry and $\bar{g}_I$ standing for \enquote{dimensionful} couplings. By expanding the right-hand side of Eq.~\eqref{mmfe5} in terms of the same basis $\left\{\mathcal{O}^{I}\right\}$, it is possible to project out the flow of each coupling $\bar{g}_I$, i.e., to compute the beta functions of the theory. In standard local field-theoretic settings, one is interested in the dimensionless version of the couplings and their flow as these contain the information on (quantum) scale invariant points.
In the present case, there is no local notion of scale; nevertheless \enquote{dimensionless} couplings can be defined that absorb factors of $N$. Specificially, the coupling $\bar{g}_I$ is related to its dimensionless counterpart $g_I$ by the relation $\bar{g}_I = N^{[d_I]}g_I$, where $[d_I]$ denotes the canonical scaling dimension of $\bar{g}_I$. Since the renormalization group parameter $N$ is dimensionless, the assignment of canonical dimensions to couplings does not follow from simple dimensional analysis. Actually, as discussed in~\cite{Eichhorn:2014xaa,Eichhorn:2017xhy,Eichhorn:2018ylk}, the scaling with $N$ is fixed by demanding that, at large-$N$, the system of beta functions is non-trivial and autonomous, i.e., the flow does not depend explicitly on $N$. One can follow similar arguments in the framework of Dyson-Schwinger equations, see, e.g.,~\cite{Pascalie:2018nrs} as well as the Polchinski equation~\cite{Krajewski:2015clk}.
The beta functions of the dimensionless couplings are thus
\begin{equation}
\partial_t g_I = -[d_I]g_I+N^{-[d_I]}\partial_t \bar{g}_I\,.
\label{mmfe7}
\end{equation}
The function $\partial_t \bar{g}_I$ is read off from the flow equation \eqref{mmfe5} by a suitable projection onto the basis defined by Eq.~\eqref{mmfe6}.

In this work, we are interested in multi-matrix models. As in standard field theory with multiple fields, the flow equation can easily be derived in this setting. For concreteness, we consider a model with two interacting Hermitian matrices $A$ and $B$ (but the argument extends to a generic set of matrices) with the following partition function,
\begin{equation}
Z[J_A,J_B] = \int\mathrm{d}A\mathrm{d}B\,\mathrm{e}^{-S [A,B]+\Tr\left(J_A A\right) + \Tr\left(J_B B\right)}\,,
\label{mmfe8}
\end{equation}
with $J_A$ and $J_B$ being the sources associated, respectively, to $A$ and $B$ and $S[A,B]$ is the classical action of the model. The regulator $\Delta S_N$ is introduced as
\begin{equation}
\Delta S_N = \frac{1}{2}\sum_{I,J}\sum_{a,b,c,d}\Phi^{I}_{ab}R^{ab,cd}_{N,IJ}(a,b)\Phi^{J}_{cd}\,,
\label{mmfe9}
\end{equation}
where $\Phi^{I}\equiv\left\{A,B\right\}$ and $R^{ab,cd}_{N,IJ}(a,b)$ satisfies the properties~\eqref{Rproperties1},~\eqref{Rproperties2} and~\eqref{Rproperties3} transferred to the multi-matrix case. Thence, the flow equation is derived in full analogy to the single-matrix model, leading to
\begin{equation}\label{flowequationMMM}
\partial_t \Gamma_N = \frac{1}{2}\mathrm{Tr}\left[\sum_{I,J}\left(\frac{\delta^2\Gamma_N}{\delta\Phi\delta\Phi}+R_{N}\right)^{-1}_{IJ}\partial_t R_{N,JI}\right]\,,
\end{equation}
where $\mathrm{Tr}$ represents a sum over matrix indices. Equation~\eqref{flowequationMMM} easily extends to a generic number of matrices. In the next subsections, we will explicitly apply it to the case of the two-matrix model related to CDT in $2+1$ dimensions, the details of which we specified above.\\

As it is crucial for the developments that follow, we emphasize an important property of the flow equation: While the path integral Eq.~\eqref{mmfe8} requires the specification of a classical action $S$, the flow equation is independent of the classical action. Instead, it provides a local vector field in the space of all couplings, indicating how the dynamics changes under a finite RG step. A classical action can be specified as an initial condition to integrate this flow and obtain the effective action. On the other hand, the flow equation can also be used to search for special points in the space of couplings, which correspond to fixed points of the RG flow. Such fixed points then give rise to a particular proposal for a classical (or microscopic) action as the starting point of the flow.

\subsection{General setup of the flow equations}\label{Sec:generalsetup}

To obtain the set of beta functions of the couplings, one has to project the flowing action onto the couplings of the corresponding operators. In practice, this is achieved by means of a series expansion of the rhs of Eq.~\eqref{flowequationMMM} in terms of powers of $A$ and $B$, the $\mathcal{P}^{-1}\mathcal{F}$ expansion. To this end, one rewrites the denominator of the flow equation as
\begin{equation}
\Gamma_N^{(2)}+R_N=\mathcal{P}_N+\mathcal{F}_N
\end{equation}
with the fluctuation matrix $\mathcal{F}_N$ and inverse propagator defined by 
\begin{equation}\label{inversepropagator}
\mathcal{P}_N = \left(\Gamma_N^{(2)}+R_N\right)\Big|_{A=B=0},\qquad \mathrm {\rm and} \qquad  \mathcal{F}_N = \Gamma_N^{(2)}-\left(\Gamma_N^{(2)}\Big|_{A=B=0}\right)\,.
\end{equation}
The right-hand side of the flow equation is then expanded as
\begin{align}
\partial_t\Gamma_N &= \frac{1}{2}\mathrm{STr}\left[ \left(\Gamma_N^{(2)}+R_N\right)^{-1}\partial_t R_N\right]\nonumber\\ &=\frac{1}{2} \mathrm{STr}\left[\left(\sum_{n=0}^{+\infty}(-1)^n\left(\mathcal{P}_N^{-1}\mathcal{F}_N\right)^n\right)\mathcal{P}_N^{-1}\,\partial_t R_N\right],
\end{align}
wherein $\mathrm{STr}$ denotes a sum over the available matrices and the indices thereof. In the next step, we split the Hermitian matrices $A$ and $B$ into symmetric and anti-symmetric parts
\begin{align}
A_{ab}=A_{1ab}+i A_{2ab},\qquad \mathrm{and}\qquad B_{ab}=B_{1ab}+i B_{2ab}\,,
\end{align}
where $A_{1ab}= A_{1ba}$, $A_{2ab}= -A_{2ba}$, $B_{1ab}= B_{1ba}$ and $B_{2ab}= -B_{2ba}$ hold. With this parameterization, we compute the variations for the Hessian. After doing that, the next step typically is to choose a particular field configuration which facilitates the projection of the right-hand side of the flow equation onto the corresponding beta function of interest.  For the purposes of this work, it suffices to take $A_{2ab}=B_{2ab}=0$. The Hessian then has the following structure, 
\begin{equation}
\left(\Gamma_N^{(2)}\right)_{ab,cd} =
\begin{pmatrix}
  \Gamma_{N, A_{1}A_{1}} & 0 & \Gamma_{N, A_{1}B_{1}} & 0\\ 
  0 & \Gamma_{N, A_{2}A_{2}} & 0 & \Gamma_{N, A_{2}B_{2}}\\ 
  \Gamma_{N, B_{1}A_{1}} & 0 & \Gamma_{N, B_{1}B_{1}} & 0\\ 
  0 & \Gamma_{N, B_{2}A_{2}} & 0 & \Gamma_{N, B_{2}B_{2}}
\end{pmatrix}_{ab,cd}.
\end{equation}
In the very last step, after having computed the beta functions which also map combinatorial differences between operators of the same power in $A$ and $B$, we project the remaining field configurations onto $A_{1ab}=a_1 \delta_{ab}$ and $B_{1ab}=b_1 \delta_{ab}$, see Sec.~\ref{sec:betafunctionsfps}.

The regulator is chosen in a diagonal form
\begin{equation}
\left(R_N\right)_{ab,cd} =
 \begin{pmatrix}
  R_{N, A_1 A_1} & 0 & 0 & 0\\ 
  0 & R_{N, A_2 A_2} & 0 & 0\\ 
  0 & 0 & R_{N, B_1 B_1} & 0\\ 
  0 & 0 & 0 & R_{N, B_2 B_2}
\end{pmatrix}_{ab,cd}
\end{equation}
with
\bea
  R_{N,A_1 A_1/A_2 A_2}(a,b)_{ab,cd} &=& R_{N,AA}(a,b)\left(\delta_{ac}\delta_{bd} \pm \delta_{ad}\delta_{bc}\right)\nonumber\\
  &{\equiv}&\frac{Z_{A}}{2} f_N(a,b)\left(\delta_{ac}\delta_{bd} \pm \delta_{ad}\delta_{bc}\right),
\eea
and
\bea
    R_{N, B_1 B_1/ B_2 B_2}(a,b)_{ab,cd}&=&R_{N,BB}(a,b)\left(\delta_{ac}\delta_{bd} \pm \delta_{ad}\delta_{bc}\right)\nonumber\\
    &{\equiv}&\frac{Z_{B}}{2} f_N(a,b)\left(\delta_{ac}\delta_{bd}\pm \delta_{ad}\delta_{bc}\right).
\eea
Therein, $Z_A, Z_B$ are wave-function renormalization factors and the details of the cut-off procedure are captured by the function
\begin{equation}
f_N(a,b)= \left(\frac{2N}{a+b}-1\right)\theta\left(1-\frac{a+b}{2N}\right),
\end{equation}
so that the regulator is closely modelled after Litim's cutoff~\cite{Litim:2001up}. With these expressions, the computation of the inverse propagator $\mathcal{P}_N$, Eq.~\eqref{inversepropagator}, is straightforward. A relevant detail repeatedly used in the $\mathcal{P}^{-1}\mathcal{F}$ expansion for concrete calculations is then given by the approximation
\begin{align}
    \sum_{a}\sum_{b}\left(\frac{1}{1+f_N(a,b)}\right)^n&\partial_t R_{N, AA/BB}(a,b)\nonumber\\& \sim N^2\frac{4\left(2+n-\eta_{A/B}\right)}{2+3n+n^2}\left(1+\mathcal{O}\left(1/N\right)\right),
\end{align}
valid at leading order in $1/N$. In this expression, we introduce the anomalous dimensions as $\eta_{A/B}=-\partial_t\ln Z_{A/B}$. 

Finally, as discussed in Refs.~\cite{Eichhorn:2013isa,Eichhorn:2018phj} and Sec.~\ref{sec:FRGmethodformatrices}, the couplings in matrix and tensor models have an inherent dimensionality in spite of having no natural notion of momentum- scale. It dictates their behavior with respect to rescalings in $N$. In our model, one has the following rescalings for the dimensionful couplings\footnote{In order to define the correct canonical scaling with $N$ for each coupling, one demands that the system of beta functions is autonomous, i.e., does not depend explicitly on $N$, at large $N$ and is non-trivial. Typically, this leads to upper bounds on the power of $N$ and our prescription is to saturate the bounds.}
\begin{equation}\label{rescalingalphas}
    \bar{\alpha}_{1/2}=\frac{\alpha_{1/2}Z_{A/B}^2}{N},
\end{equation}
and 
\begin{equation}\label{rescalingbeta}
    \bar{\beta}=\frac{\beta Z_A Z_B}{N}\,,
\end{equation}
where $\alpha_{1/2}$ and $\beta$ are \enquote{dimensionless} couplings. This assignment of scaling dimensionality is chosen in such a way as to facilitate a $1/N$-expansion of the beta functions where the leading coefficient is $\mathcal{O}(N^0)$. Because of this choice a sensible continuum limit exists.

Equipped with this we are ready to calculate the beta functions from the flow equation.

\subsection{Beta functions and fixed points for the $ABAB$ matrix model}\label{sec:betafunctionsfps}

In this work we base our analysis on the simplest possible truncation of the flowing action adapted from the action used in Eq.~\eqref{mmmparititionfunction}, i.e.,
\begin{equation}
 \Gamma_{N} [A,B]=\frac{Z_{A}}{2}\Tr\left(AA\right)+\frac{Z_{B}}{2}\Tr\left(BB\right)-\frac{\bar{\alpha}_1}{4}\Tr\left(AAAA\right)-\frac{\bar{\alpha}_2}{4}\Tr\left(BBBB\right)-\frac{\bar{\beta}}{2}\Tr\left(ABAB\right).
\label{truncefa}
\end{equation}
Compared to the analysis of the single-matrix model~\cite{Eichhorn:2013isa,Eichhorn:2014xaa}, we stick to the sign conventions of Refs.~\cite{Kazakov:1998qw,ZinnJustin:2003kq} in which the interaction terms carry a negative sign. There is another single-trace interaction compatible with the symmetries of the model which is $\mathrm{Tr}(ABBA)$. However, the beta function of the coupling associated to this interaction is proportional to itself in this simple truncation. Hence, such a coupling can be consistently set to zero and we do not include it here.

From Eq.~\eqref{truncefa}, one can extract the fluctuation matrix $\mathcal{F}_N$ which is the remaining ingredient we need in order to compute the $\mathcal{P}^{-1}\mathcal{F}$ expansion. Its non-vanishing entries are given by:
\bea
\left(\mathcal{F}_{A_1 A_1}\right)_{ab,cd}&=&-\frac{\bar{\beta}}{2} \left(B_{1 ad}B_{1 bc}+B_{1 ac}B_{1 bd}\right)-\frac{\bar{\alpha}_1}{2}\bigl(A_{1 ad}A_{1 bc}+A_{1 ac}A_{1 bd}+A_{1 bn}A_{1 dn}\delta_{ac}\nonumber\\
&{}&+A_{1 bn}A_{1 cn}\delta_{ad}+A_{1 an}A_{1 dn}\delta_{bc}+A_{1 an}A_{1 cn}\delta_{bd}\bigr),\\
\left(\mathcal{F}_{A_1 B_1}\right)_{ab,cd}&=&-\frac{\bar{\beta}}{2}\left(A_{1 dn} B_{1 bn}\delta_{ac}+A_{1 cn} B_{1 bn}\delta_{ad}+A_{1 dn} B_{1 an}\delta_{bc}+A_{1 cn} B_{1 an}\delta_{bd}\right),\\
\left(\mathcal{F}_{A_2 A_2}\right)_{ab,cd}&=&\frac{\bar{\beta}}{2} \left(B_{1 ad}B_{1 bc}-B_{1 ac}B_{1 bd}\right)+\frac{\bar{\alpha}_1}{2}\bigl(A_{1 ad}A_{1 bc}-A_{1 ac}A_{1 bd}-A_{1 bn}A_{1 dn}\delta_{ac}\nonumber\\
&{}&+A_{1 bn}A_{1 cn}\delta_{ad}+A_{1 an}A_{1 dn}\delta_{bc}-A_{1 an}A_{1 cn}\delta_{bd}\bigr),\\
\left(\mathcal{F}_{A_2 B_2}\right)_{ab,cd}&=&\frac{\bar{\beta}}{2}\left(-A_{1 dn} B_{1 bn}\delta_{ac}+A_{1 cn} B_{1 bn}\delta_{ad}+A_{1 dn} B_{1 an}\delta_{bc}-A_{1 cn} B_{1 an}\delta_{bd}\right),
\eea
\bea
\left(\mathcal{F}_{B_1 B_1}\right)_{ab,cd}&=&-\frac{\bar{\beta}}{2} \left(A_{1 ad}A_{1 bc}+A_{1 ac}A_{1 bd}\right)-\frac{\bar{\alpha}_2}{2}\bigl(B_{1 ad}B_{1 bc}+B_{1 ac}B_{1 bd}+B_{1 bn}B_{1 dn}\delta_{ac}\nonumber\\
&{}&+B_{1 bn}B_{1 cn}\delta_{ad}+B_{1 an}B_{1 dn}\delta_{bc}+B_{1 an}B_{1 cn}\delta_{bd}\bigr),\\
   \left(\mathcal{F}_{B_2 B_2}\right)_{ab,cd}&=&\frac{\bar{\beta}}{2} \left(A_{1 ad}A_{1 bc}-A_{1 ac}A_{1 bd}\right)+\frac{\bar{\alpha}_2}{2}\bigl(B_{1 ad}B_{1 bc}-B_{1 ac}B_{1 bd}-B_{1 bn}B_{1 dn}\delta_{ac}\nonumber\\
   &{}&+B_{1 bn}B_{1 cn}\delta_{ad}+B_{1 an}B_{1 dn}\delta_{bc}-B_{1 an}B_{1 cn}\delta_{bd}\bigr).
\eea

In the following, we present the computations relevant to obtain the set of beta functions from the flow equation. First, we briefly summarize our line of action: Given the simple truncation, we only need to compute the $\mathcal{P}^{-1}\mathcal{F}$ expansion up to order $2$. At $0$th order one obtains a field-independent constant which can be absorbed in the normalization of the path integral and is thus of no interest hereafter. From the $1$st- order- contributions, we compute the expressions for the anomalous dimensions $\eta_A$ and $\eta_B$. Then, at $2$nd order we yield the beta functions for the couplings.

Starting off with the $1$st order of the expansion, we have
\begin{align}
    \partial_t\Gamma_N\Big|_{\Tr(A^2), \Tr(B^2)} &= -\frac{1}{2}\mathrm{STr}\left[\left(\partial_t R_N\right)\left(\mathcal{P}_N^{-1}\right)\left(\mathcal{F}_N\right)\left(\mathcal{P}_N^{-1}\right)\right]\nonumber\\
    &=\sum_{a,b}\Biggl[\frac{1}{Z_A^2}\frac{\partial_t R_{N,AA}(a,b)}{(1+f_N(a,b))^2}\left(\bar{\beta}B_{1aa}B_{1bb}+\bar{\alpha}_1 A_{1aa}A_{1bb}+2\bar{\alpha}_1 A_{1 bn}A_{1 bn}\right)\nonumber\\
    &+\frac{1}{Z_B^2}\frac{\partial_t R_{N,BB}(a,b)}{(1+f_N(a,b))^2}\left(\bar{\beta}A_{1aa}A_{1bb}+\bar{\alpha}_2 B_{1aa}B_{1bb}+2\bar{\alpha}_2 B_{1 bn}B_{1 bn}\right)\Biggr].
\end{align}
In the next step we focus only on single-trace contributions and thus discard any occurring double-trace terms (which have the structure $\mathrm{Tr}(\ldots)\mathrm{Tr}(\ldots)$). Then, as discussed in Sec.~\ref{Sec:generalsetup}, the second part of the projection is applied, namely inserting $A_{1ab}=a_1 \delta_{ab}$ and $B_{1ab}=b_1 \delta_{ab}$ and carrying out the summations. At large $N$, this leads to
\begin{align}
&\left(N\partial_N\frac{Z_A}{2}\right)(N a_1^2)+\left(N\partial_N\frac{Z_B}{2}\right)(N b_1^2)\nonumber\\ &=\frac{1}{Z_A^2 Z_B^2}\left[Z_B^2 Z_A \bar{\alpha}_1 a_1^2 N^2 \frac{(4-\eta_A)}{3}+Z_A^2 Z_B \bar{\alpha}_2 b_1^2 N^2 \frac{(4-\eta_B)}{3}\right]
\nonumber\\&=Z_A a_1^2 N \alpha_1\frac{(4-\eta_A)}{3}+Z_B a_2^2 N \alpha_2\frac{(4-\eta_B)}{3}.
\end{align}
In the last step we used the rescaling of the dimensionful couplings $\bar{\alpha}_{1/2}$, as given in Eq.~\eqref{rescalingalphas}. By comparison with the left-hand side, we extract
the anomalous dimensions
\begin{equation}\label{eq:anomdim}
    \eta_{A/B}=\frac{8\alpha_{1/2}}{-3+2\alpha_{1/2}}\,,
\end{equation}
 which results from the solution of an algebraic equation.
 
At $2$nd order of the expansion, we obtain
\begin{align}
    &\partial_t\Gamma_N\Big|_{\Tr(A^4), \Tr(B^4), \Tr((AB)^2)}\nonumber\\&=\frac{1}{2}\mathrm{STr}\left[\left(\partial_t R_N\right)\left(\mathcal{P}_N^{-1}\right)\left(\mathcal{F}_N\right)\left(\mathcal{P}_N^{-1}\right)\left(\mathcal{F}_N\right)\left(\mathcal{P}_N^{-1}\right)\right]\nonumber\\
    &=\frac{1}{2 Z_A^3 Z_B^3}\sum_{a,b}\Biggl[\partial_t R_{N,AA}(a,b)\left\{\bar{\alpha}_1^2 Z_B^3 S_1 + \bar{\alpha}_1\bar{\beta} Z_B^3 S_2 + \bar{\beta}^2 Z_A Z_B^2 S_3 + \bar{\beta}^2 Z_B^3 S_4\right\}\nonumber\\&\qquad+\partial_t R_{N,BB}(a,b)\bigl\{\bar{\alpha}_2^2 Z_A^3 S_1(A\leftrightarrow B) + \bar{\alpha}_2\bar{\beta} Z_A^3 S_2(A\leftrightarrow B) \nonumber\\&\qquad+ \bar{\beta}^2 Z_B Z_A^2 S_3(A\leftrightarrow B) + \bar{\beta}^2 Z_A^3 S_4(A\leftrightarrow B)\bigr\}\Biggr]\,, 
\end{align}
where $S_1$ to $S_4$ are given by
\begin{align}
    S_1 &=8\sum_h\frac{A_{1aa} A_{1bn} A_{1hn} A_{1hb}}{(1+f_N(a,b))^2(1+f_N(a,h))}+6\sum_{g,h}\frac{A_{1ah} A_{1ha}A_{1bg}A_{1gb}}{(1+f_N(a,b))^2(1+f_N(g,h))}\nonumber\\&\qquad+4\sum_{g}\frac{A_{1an}A_{1gn}A_{1gm}A_{1ma}}{(1+f_N(a,b))^2(1+f_N(g,b))},\nonumber\\
    S_2 &=  8\sum_{h}\frac{B_{1aa} B_{1hb}B_{1bn}A_{1hn}}{(1+f_N(a,b))^2(1+f_N(a,h))}+4\sum_{g,h}\frac{A_{1ha}B_{1ha}A_{1gb}B_{1gb}}{(1+f_N(a,b))^2(1+f_N(g,h))},\nonumber\\
    S_3 &= 4\frac{A_{1bn}B_{1bn}A_{1am}B_{1am}}{(1+f_N(a,b))^3}+4\sum_{h}\frac{A_{1hn}A_{1bm}B_{1bn}B_{1hm}}{(1+f_N(a,b))^2(1+f_N(a,h))},\nonumber\\
    S_4 &=2 \sum_{g,h}\frac{B_{1ah}B_{1ah}B_{1bg}B_{1bg}}{(1+f_N(a,b))^2(1+f_N(g,h))}.
\end{align}
Again focusing on the single-trace terms only and  setting $A_{1ab}=a_1 \delta_{ab}$ and $B_{1ab}=b_1 \delta_{ab}$, at large $N$ one finds
\begin{align}
    &-\left(N\partial_N \frac{\bar{\alpha}_1}{4}\right)(N a_1^4)
    -\left(N\partial_N \frac{\bar{\alpha}_2}{4}\right)(N b_1^4)
    -\left(N\partial_N \frac{\bar{\alpha}_2}{2}\right)(N a_1 b_1 a_1 b_1)\nonumber\\&=\frac{1}{2Z_A^3 Z_B^3}\Big[\frac{Z_A}{10}(5-\eta_A) N^2\left(\bar{\alpha}_1^2 Z_B^3 4 a_1^4+\bar{\beta}^2 Z_A Z_B^2 4 (a_1b_1a_1b_1)\right)\nonumber\\&\qquad+\frac{Z_B}{10}(5-\eta_B) N^2\left(\bar{\alpha}_2^2 Z_A^3 4 b_1^4+\bar{\beta}^2 Z_B Z_A^2 4 (a_1b_1a_1b_1)\right)\Big]\nonumber\\&=\frac{Z_A^2}{5}(5-\eta_A)\alpha_1^2 a_1^4+\frac{Z_A Z_B}{5}(5-\eta_A)\beta^2 (a_1 b_1 a_1 b_2)\nonumber\\&\qquad+\frac{Z_B^2}{5}(5-\eta_B)\alpha_2^2 b_1^4+\frac{Z_A Z_B}{5}(5-\eta_B)\beta^2 (a_1 b_1 a_1 b_2).
\end{align}
In the last step we used the rescaling of the dimensionful couplings $\bar{\alpha}_1,\bar{\alpha}_2$ and $\bar{\beta}$, see Eqs.~\eqref{rescalingalphas} and~\eqref{rescalingbeta}.

We therefore obtain the following  beta functions,
\begin{eqnarray}
\beta_{\alpha_1}&=&\left(2\eta_A +1 \right)\alpha_1 -\frac{4}{5}\left(5-\eta_A \right)\alpha_1^2,\\
\beta_{\alpha_2}&=& \left(2\eta_B +1 \right)\alpha_2 -\frac{4}{5}\left(5-\eta_B \right)\alpha_2^2,\\
\beta_{\beta}&=&\left(\eta_A+\eta_B+1 \right)\beta-\frac{2}{5}\left( (5-\eta_A)+ (5-\eta_B)\right)\beta^2.
\end{eqnarray}
All three  beta functions reflect the exchange symmetry $A \leftrightarrow B$. The same symmetry determines the fixed-point structure: A fixed point must either be symmetric under the exchange $\alpha_1 \leftrightarrow \alpha_2$, or come with a counterpart such that the pair of fixed points can be mapped into each other under this exchange. Furthermore, for $\alpha_1=\alpha_2=\beta$ the system of beta functions is governed by a single equation, reflecting the fact that the model then exhibits a hidden $\text{U}(1)$ symmetry which is otherwise broken~\cite{zinnjustin1999matrix,Kostov:1988fy,Gaudin:1989vx,Kostov:1992pn,Eynard:1992cn,Eynard:1995nv}.
\\
Comparing with previous work, we find agreement with the  beta function for the single-matrix quartic coupling reported in~\cite{Eichhorn:2013isa}, as expected (note the difference in sign of the quartic term in the truncation).

As a side remark, we mention that when extending the single-trace truncation used here by the next higher-order term, no contribution of type $\Tr\left((AB)^3\right)$ is included in the effective action since it would violate the $\mathbb{Z}_2$-symmetry of the model. Enlargening by $\Tr\left(A^6\right)$ and $\Tr\left(B^6\right)$, only the  beta functions for $\alpha_1$ and $\alpha_2$ receive further contributions in agreement with the single-matrix model limit~\cite{Eichhorn:2013isa}.

\subsubsection{Fixed-point structure in the symmetric limit}\label{section:symmetriclimit}
 
 The system analyzed above lends itself to an enhancement of the symmetry to an $A \leftrightarrow B$ exchange symmetry, which entails a smaller theory space since it requires setting $\alpha_1 = \alpha_2\equiv\alpha$, as well as $\eta_A = \eta_B$. In addition, when also $\alpha=\beta$ holds, the symmetry of the system is enhanced even further, since the model then displays a $\text{U}(1)$ invariance where the matrices $A$ and $B$ behave as a vector under $\text{U}(1)$ transformations. In these symmetry-enhanced cases we obtain the fixed-point structure in Tab.~\ref{table:1}, where we use the convention that the critical exponents $\theta_i$ are the eigenvalues of the stability matrix, multiplied by an additional negative sign, i.e.,
\be
\theta_i = - {\rm eig} \left(\frac{\partial \beta_{g_a}}{\partial g_b} \right)\Big|_{\vec{g}= \vec{g}_{\ast}}.\label{eq:theta}
\ee
Herein, $\vec{g}$ denotes the vector of all couplings.

\begin{table}[ht!]
\centering
\begin{tabular}{||c||c || c c || c c|| c ||} 
 \hline
  Fixed point & type of FP & $\alpha_\ast$ & $\beta_\ast$ & $\theta_1$ & $\theta_2$ & $\eta$ \\ [0.5ex] 
 \hline\hline\hline
 A & GFP & 0 & 0 & -1 & -1 & 0 \\ 
 \hline
 B & NGFP (SP) & 0 & 0.25 & -1 & 1 & 0\\
 \hline
 D & NGFP (SP) & 0.10 & 0 & 1.07 & -0.43 & -0.29\\
 \hline
 C & NGFP & 0.10 & 0.10 & 1.07 & 0.43 & -0.29\\
 \hline
\end{tabular}
\caption{Fixed-point values of the couplings and respective critical exponents in the symmetric limit where $\alpha_1=\alpha_2\equiv \alpha$ holds. By SP (saddle point) we denote fixed points with one relevant direction, which can act as either IR or UV fixed points of the flow.}
\label{table:1}
\end{table}

For the $\text{U}(1)$ symmetric fixed point C, the first critical exponent, $\theta_1=1.07$, is also recovered within the $\text{U}(1)$ symmetric theory space, where only a single quartic interaction exists. The second critical exponent, $\theta_2=0.43$, encodes the relevance of $\text{U}(1)$-symmetry-breaking perturbations. Accordingly, we conclude that the enhancement of the symmetry in the IR requires tuning, as the fixed point is not IR attractive from outside the $\text{U}(1)$ symmetric theory space.

\begin{figure}[!t]
\includegraphics[angle=0,width=9cm]{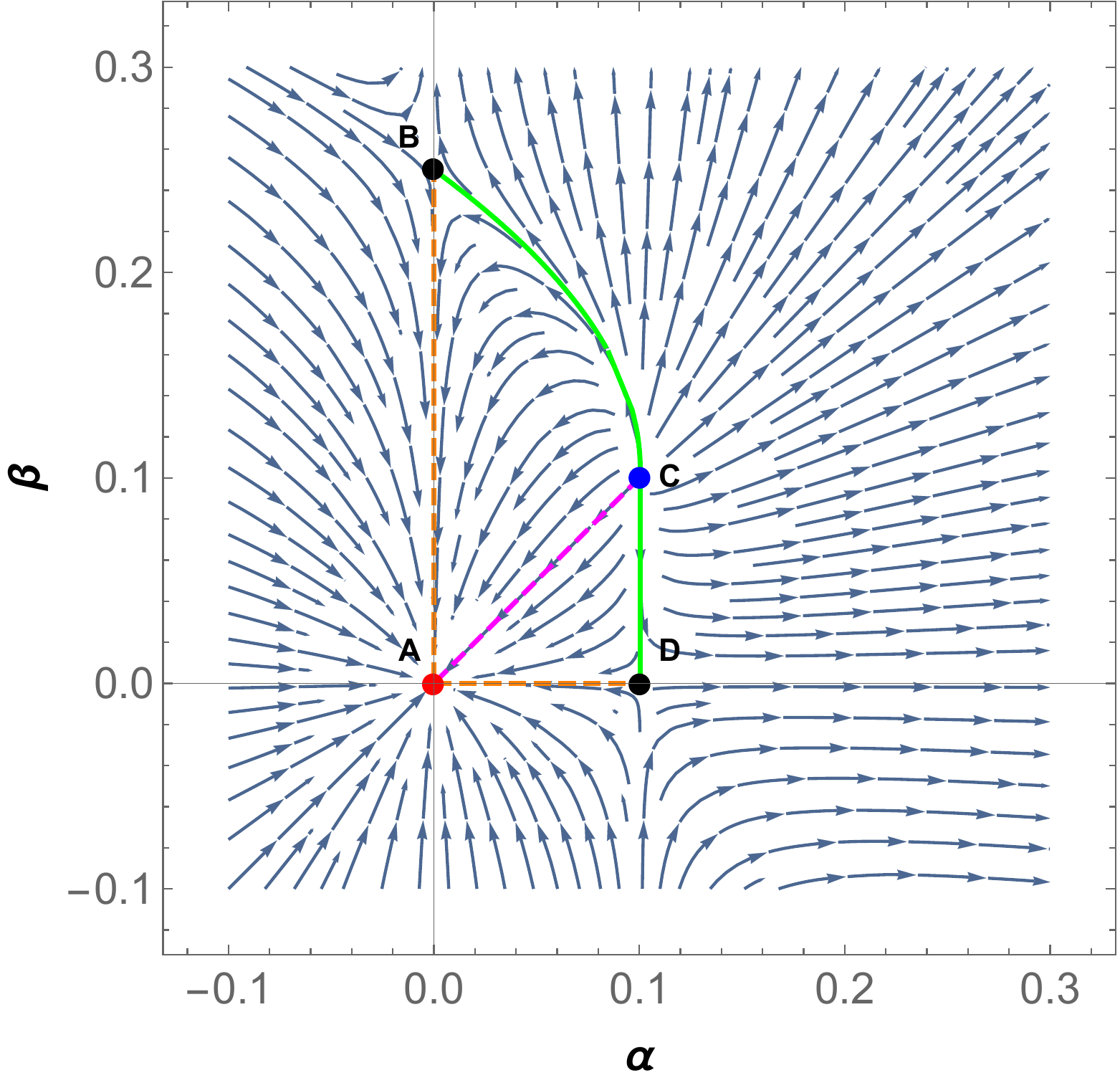}
\centering
\caption{Phase diagram of the $ABAB$ matrix model with $\alpha_1=\alpha_2=\alpha$ where $4$ fixed points are shown. The black dots ($B,D$) denote saddle points, the red dot ($A$) corresponds to the GFP while the blue dot ($C$) marks the NGFP of interest. For the dashed pink line one has $\alpha=\beta$. Arrows point towards the IR, i.e., small $N$.}\label{figure:phase_diagram}
\end{figure}

At $\alpha = 1.5$, we observe a singularity of the flow  arising from the non-polynomial structure for the anomalous dimension, cf.~Eq.~\eqref{eq:anomdim} beyond which we have discarded any further  zeros of the beta functions.  These are characterized by anomalous dimensions beyond the regulator bound $\eta_{\rm max}=1$, which is required for a well-defined regulator, see~\cite{Meibohm:2015twa,Eichhorn:2018ylk} for more details. Further, they exhibit very large critical exponents $\mathcal{O}(10)$, and therefore fail the a-posteriori-check of the self-consistency of our truncation. The latter is based on the assumption of near-canonical scaling, and therefore requires deviations of the critical exponents from the canonical dimensions to be $\mathcal{O}(1)$.
The  fixed-point candidates are illustrated in Fig.~\ref{figure:phase_diagram}. The critical exponents for these are compatible with our assumption of near-canonical scaling. Further, the anomalous dimension $\eta$ is relatively small. Accordingly, we only find changes of the fixed-point properties at the percent level when we compare to the perturbative approximation, where the $\eta$s arising from loop factors are neglected.

\subsubsection{Fixed-point structure in the asymmetric case}\label{section:asymmetriclimit}
The more general theory space for the $ABAB$ matrix model does not require the exchange symmetry between $A$ and $B$ to hold at each point during the flow. Accordingly, there can even be fixed points outside the symmetry-enhanced theory space in which $\alpha_1=\alpha_2$. Such fixed points appear in pairs which are mapped into each other under the mapping $A \leftrightarrow B$. Further, the symmetry-enhanced fixed points from Tab.~\ref{table:1} are also necessarily fixed points of the more general flow. Within the larger theory space, they can be IR attractive or repulsive, i.e., symmetry enhancement can occur as a natural consequence of the flow. In Tab.~\ref{table:2}, we list fixed-point candidates and their properties. We highlight the symmetry-enhanced fixed points from Tab.~\ref{table:1} in italics; the additional critical exponent which indicates whether the symmetry-enhancement is an automatic consequence of the flow (which requires $\theta<0$) is indicated in bold.

\begin{table}[ht!]
\centering
\begin{tabular}{||c ||c || c c c || c c c|| c c||} 
 \hline
 Fixed point & type of FP & $\alpha_{1\, \ast}$ & $\alpha_{2\, \ast}$ & $\beta_\ast$ & $\theta_1$ & $\theta_2$ & $\theta_3$ & $\eta_A$ & $\eta_B$\\ 
 \hline\hline\hline
 A & \emph{GFP} & 0 & 0 & 0 & -1 & -1 & $\mathbf{-1}$ & 0 &0 \\ 
 \hline
 D & NGFP (SP)& 0 & 0 & 0.25 & 1& -1 & -1  & 0 & 0\\
 \hline
 E & NGFP (SP) & 0 & 0.10 & 0.17 & 1.07 & 0.71 & -1 & 0 & -0.29\\
 \hline
 F & NGFP (SP) & 0 & 0.10 & 0 & 1.07 & -0.71 & -1 & 0 & -0.29\\
 \hline
 C & NGFP (SP) & 0.10 & 0 & 0.17 & 1.07 & 0.71 & -1 & -0.29 & 0\\ 
 \hline
B & NGFP (SP) & 0.10 & 0 & 0 & 1.07 & -0.71 & -1 & -0.29 & 0\\
 \hline
 H & \emph{NGFP} & 0.10 & 0.10 & 0.10 & 1.07 &  $\mathbf{1.07}$& 0.43 &-0.29 & -0.29\\
 \hline
 G & \emph{NGFP (SP)} & 0.10 & 0.10 & 0 & 1.07 & $\mathbf{1.07}$ & -0.43 &-0.29 & -0.29 \\
  \hline
\end{tabular}
\caption{Fixed-point values of the couplings and respective critical exponents  and anomalous dimensions in the case of the asymmetric model where in general $\alpha_1\neq\alpha_2\neq \beta$ holds.  Fixed points which exist in the symmetry-enhanced theory space with $\alpha_1=\alpha_2$ are indicated in italics, cf.~Tab.~\ref{table:1}. The additional critical exponent that such fixed points have in the larger theory space are indicated in bold.}
\label{table:2}
\end{table}

As expected, fixed points with $\alpha_1\neq \alpha_2$ come in pairs which map into each other under $\alpha_1 \leftrightarrow \alpha_2$, cf.~third and fifth as well as fourth and sixth line of Tab.~\ref{table:2}. It is interesting to observe that at these fixed points, one of the two sectors features no self-interactions, and can therefore be integrated out exactly in the path integral.

\begin{figure}[!t]
\includegraphics[angle=0,width=11cm]{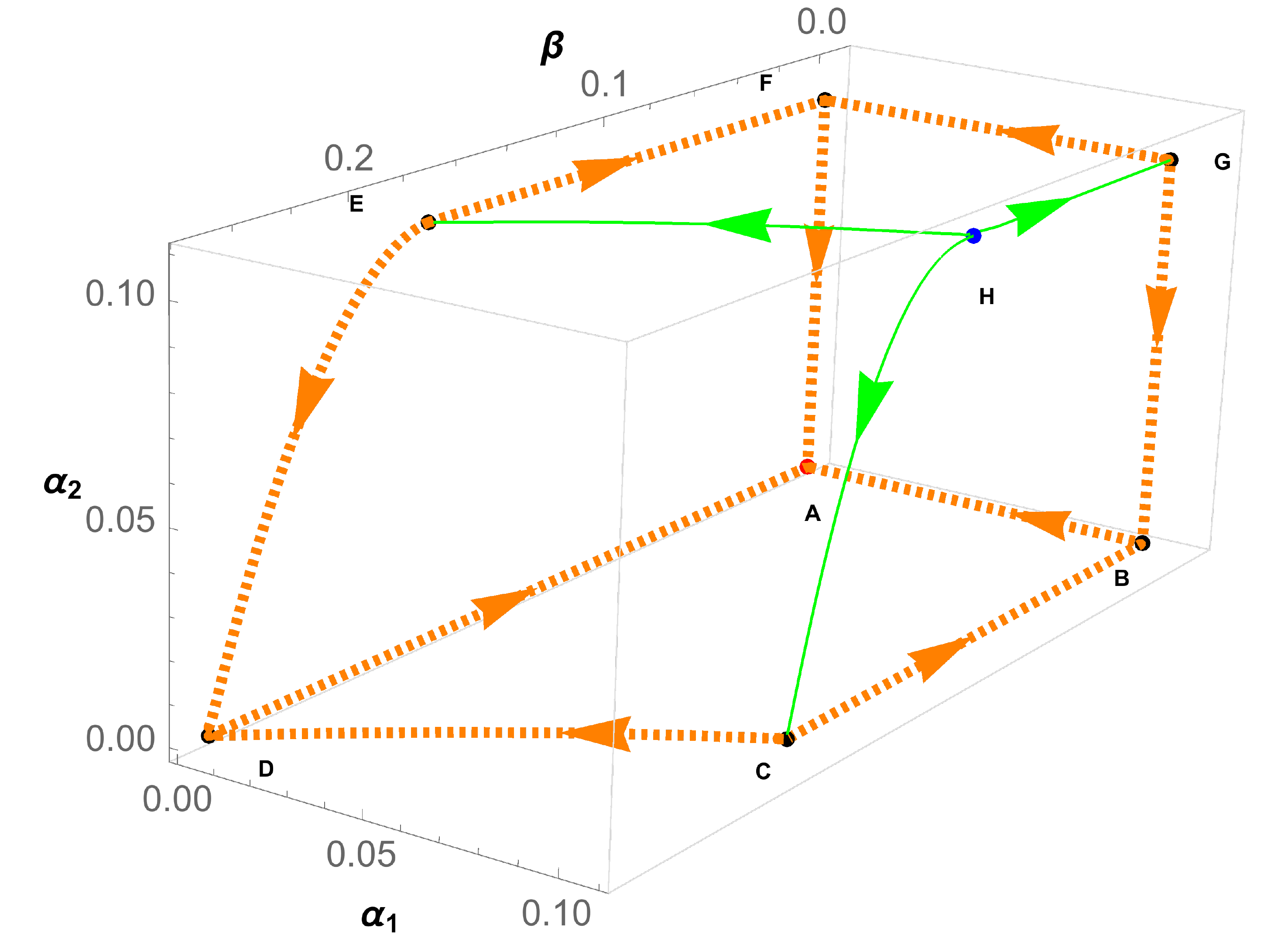}
\centering
\caption{Phase diagram of the $ABAB$ matrix model with $\alpha_1\neq\alpha_2$ where $8$ fixed points are shown. The black dots ($B,C,D,E,F,G$) denote saddle points, the red dot ($A$) corresponds to the GFP while the blue dot ($H$) marks the NGFP of interest. Arrows point towards the IR, i.e., small $N$.}\label{figure:phase_diagram3d}
\end{figure}

An inspection of Tab.~\ref{table:2} highlights that symmetry-enhancement requires fine-tuning, as the additional critical exponent that characterizes the flow of the symmetry-enhanced fixed points in the larger theory space is positive for both cases. Accordingly, the surface $\alpha_1=\alpha_2$ is likely to be an IR repulsive surface, since both fixed points inside it are. Given that these two fixed points differ by one in the number of their relevant directions, there is a separatrix linking the two, providing a complete trajectory within the symmetry-enhanced fixed point that links a nontrivial universality class in the UV to a nontrivial universality class in the IR. In addition, for $\alpha_1=\alpha_2=\beta$ the symmetry of the system is further enhanced. The corresponding surface is strongly IR repulsive since the fixed point inside it has three relevant directions. Thus any quartic perturbation away from this symmetry enhancement grows under the RG flow. The fixed points and separatrices which serve to divide the space of microscopic couplings into distinct phases, are shown in Fig.~\ref{figure:phase_diagram3d}.

As expected, we recover $\theta=1.07$ for the fixed point with $\alpha_{1\, \ast}=0=\beta_{\ast}$ (and $\alpha_{2\, \ast}=0=\beta_{\ast}$, respectively). Both of those correspond to the interacting fixed point in the single Hermitian matrix model. In the single-trace approximation, the critical exponent converges to $\theta=1$ from above. As discovered in~\cite{Eichhorn:2018phj}, an alternative prescription for the critical exponents, in which $\eta=\rm const$ is implemented, improves this result to $\theta \rightarrow 0.91$, whereas the exact result, corresponding to the double-scaling limit, is\footnote{From the standard expression which defines the double-scaling limit $N(g-g_c)^{(2-\gamma_{\mathrm{str}})/2} = \mathrm{const.}$, with $g$ denoting the matrix-model coupling and $g_c$ the coupling at criticality, one can read off the relation between the critical exponent $\theta$ and the string susceptibility $\gamma_{\mathrm{str}}$, which is $\theta = 2/(2-\gamma_{\mathrm{str}})$, see for instance~\cite{Eichhorn:2013isa}.} $\theta=0.8$. In the present paper, we work with a simple truncation and do not aim at accuracy in the scaling spectrum, therefore we only provide the critical exponents evaluated according to Eq.~\eqref{eq:theta}. 

\section{Review of the phase diagram of the $ABAB$ matrix model}\label{sec:reviewpd}

The two-matrix model with $ABAB$ interaction has been exactly solved in the large $N$ limit~\cite{Kazakov:1998qw,ZinnJustin:2003kq} using the character expansion method~\cite{DiFrancesco:1992cn,Kazakov:1995ae,Kazakov:1995gm,Kazakov:1996zm,Kostov:1997bn}. Its phase diagram has been analyzed there and in particular in relation to CDT in $2+1$ dimensions in~\cite{Ambjorn:2001br,Ambjorn:2003sr,Ambjorn:2003ct}. Indeed, our results look strikingly similar to theirs, constituting a strong test of the functional RG technique. In particular, we emphasize that we obtain our phase diagrams in the simplest possible truncation of the flowing action. Thus the qualitative agreement is a very promising sign for the performance of the method.\\
Regarding quantitative results, we can compare, e.g., the critical exponents at the $\alpha_1=0=\beta$ fixed point to that of the single-matrix model. As in studies of the single-matrix model in~\cite{Eichhorn:2013isa}, the value for the relevant critical exponent deviates from the exact result corresponding to the string susceptibility $\gamma_{\text{str}}=-\frac{1}{2}$ by about 34 \%. At the $\alpha_1=0=\alpha_2$ fixed point the relevant critical exponent differs by about 17 \% from the exact result $\gamma_{\text{str}}=\frac{1}{3}$~\cite{Kazakov:1998qw}. For fixed points $C$ and $E$ in Tab.~\ref{table:2} the relevant critical exponent is off by 21 \% compared to $\gamma_{\text{str}}=-\frac{1}{3}$~\cite{Kazakov:1986hu,ZinnJustin:2003kq}. Finally, at the tri-critical point corresponding to the fixed point $H$ in Tab.~\ref{table:2}, our result departs by 7 \% from the exact value $\gamma_{\text{str}}=0$~\cite{Kazakov:1998qw,ZinnJustin:1999wt,Kazakov:1988ch}. Therefore, an extended truncation is called for to obtain quantitative results for the various universality classes. Additionally, it is an important open question what the impact of higher-order operators, in particular multi-trace ones, is on the full structure of the phase diagram.

Regarding the interpretation of the various phases, we follow the literature, in particular~\cite{Ambjorn:2001br, Kazakov:1998qw}. The different phases can be distinguished by exploring the behavior of order parameters. For instance, the scaling behavior of the loop average serves as an order parameter which allows to relate the length $L$ of a loop of $A/B$-type of links to the enclosed area $\mathcal{A}$. To make similar arguments solely using FRG technology, the composite-operator renormalization construction~\cite{Pawlowski:2005xe,Igarashi:2009tj} developed in the context of the Asymptotic Safety program in~\cite{Pagani:2016dof,Becker:2018quq,Houthoff:2020zqy}, would have to be carried over to the application of the FRG in the discrete quantum-gravity context. 

For the symmetric model ($\alpha_1=\alpha_2$) with $ABAB$ interaction term, it is possible to identify two distinct continuum phases. In~\cite{Ambjorn:2001br, Kazakov:1998qw}, the first phase is identified with the line which connects the points $D$ and $C$ in Fig.~\ref{figure:phase_diagram}, which is a separatrix in our RG flow. Indeed, initial conditions for the RG flow along that lines are the only ones that lead to the fixed point $D$ as the IR fixed point. At this fixed point, $\beta=0$, which means that one faces two decoupled single-matrix models. A fixed point at finite quartic coupling with one relevant direction is well-known to encode the double-scaling limit~\cite{Brezin:1992yc, Eichhorn:2013isa}. Thus, 
its geometric properties match those of $2d$ continuum Liouville quantum gravity~\cite{Polyakov:1981rd,DiFrancesco:1993cyw,david2016liouville} which is a model describing a $c=0$ conformal field theory coupled to $2d$ Euclidean quantum gravity, with Hausdorff dimension $d_\text{H}=4$ and the scaling relation $\mathcal{A}\sim L^2$~\cite{Kazakov:1998qw}, where $\mathcal{A}$ is the area of a closed loop and $L$ a characteristic length scale of the loop.

The second phase in the symmetric model ($\alpha_1=\alpha_2$), is identified in~\cite{Ambjorn:2001br, Kazakov:1998qw} with the line which connects points $B$ and $C$. In our RG flow, this line again corresponds to a separatrix and is the unique RG trajectory that ends in fixed point $B$. For this phase, the geometric properties are not the same as of the other phase in the symmetric model. In spite of also having $c=0$, the critical behavior is different and one finds the anomalous scaling $\mathcal{A}\sim L^{\frac{4}{3}}$, which is that of a branched polymer~\cite{Jonsson:1997gk,Ambjorn:1997jf}. These results  indicate that touching interactions are abundant in the second phase and lead to a fractal structure of the emergent geometry.

Further, we point out that both phases are separated by a phase transition which is marked by the point $C$ in Fig.~\ref{figure:phase_diagram}. As noted in~\cite{Ambjorn:2001br}, it is not clear if and how this point can be given an interpretation in terms of three-dimensional geometry and CDT in particular. However, as stated in~\cite{ZinnJustin:2003kq,Kazakov:1998qw,ZinnJustin:1999wt,Kostov:1999qx}, it can be related to a conformal field theory with central charge $c = 1$ (i.e., a free and massless boson) coupled to $2d$ Euclidean gravity. 

Finally, fixed point $A$ is the free fixed point and therefore only exhibits trivial scaling behavior.

One obtains a somewhat richer phase structure for the general case $\alpha_1 \neq \alpha_2$, see Sec.~\ref{section:asymmetriclimit}. 
Its geometric properties are discussed in~\cite{ZinnJustin:2003kq,Ambjorn:2003ct}. As far as the relation to CDT in $2+1$ dimensions is concerned, however, only the critical line for which $\alpha_1=\alpha_2$ holds (corresponding in Fig.~\ref{figure:phase_diagram3d} to the line which connects the points $H$ and $G$) is relevant~\cite{Ambjorn:2003ct}. Based on this, it has been suggested~\cite{Ambjorn:2003ct} to identify this phase of the matrix model with that of CDT in dimensions $2+1$~\cite{Ambjorn:2000dv,Ambjorn:2001cv} which, as shown by numerical evidence~\cite{Ambjorn:2000dja}, corresponds to extended three-dimensional spacetimes with Hausdorff dimension $d_\text{H}=4$ \textit{for the spatial slices}, in line with the value found for $2d$ Euclidean gravity. 
Notice, however, that without a more detailed grasp of the transfer matrix and thus of the Hamiltonian operator, it is not possible to more precisely characterize the interactions between the adjacent spatial slices and to understand how it analytically encodes the time evolution and thus the generation of extended three-dimensional geometries, so far only numerically observed by means of Monte-Carlo simulations in CDT~\cite{Ambjorn:2000dja,Ambjorn:2012jv}.

\section{Discussion and outlook} \label{sec:discussion}
In this article, we have pursued the goal to further develop functional Renormalization Group techniques for matrix and tensor models, with a particular focus on multi-field models that could constitute the key to incorporate causality into the setting. Our leading-order FRG analysis has resulted in a phase diagram in good agreement with the results presented in~\cite{Kazakov:1998qw}, providing a strong test of the functional RG approach to matrix/tensor models and highlighting that already at very low truncation order, key physics aspects are well captured.\\
To quantitatively characterize the universality classes requires extended truncations, as is obvious, e.g., from a comparison of the 2d Euclidean gravity scaling exponent that we recover in a limiting case which differs from the exact result by 34 \%. In the future, a more accurate determination of scaling exponents will allow to bridge the gap to continuum techniques to understand whether the Reuter universality class, in particular in continuum studies with Euclidean signature~\cite{Litim:2003vp,Codello:2008vh,Percacci:2010yk,Rechenberger:2012pm,Demmel:2012ub,Demmel:2013myx,Demmel:2014sga,Ohta:2012vb,Ohta:2013uca} or with foliation structure~\cite{Rechenberger:2012dt,Biemans:2016rvp,Biemans:2017zca,Houthoff:2017oam,Knorr:2018fdu}, can be recovered from the matrix/tensor model setting. At the same time, this would also allow to understand whether another gravitational universality class, that related to the tentative asymptotically free fixed point in Ho\v{r}ava-Lifshitz gravity~\cite{Horava:2009uw,Benedetti:2013pya,Contillo:2013fua,DOdorico:2014tyh,DOdorico:2015pil,Barvinsky:2015kil,Barvinsky:2017kob,Barvinsky:2019rwn}, can be recovered from the present setting, see~\cite{Horava:2009if,Benedetti:2009ge,Ambjorn:2010hu,Anderson:2011bj,Sotiriou:2011mu,Budd:2011zm,Ambjorn:2013joa,Benedetti:2014dra,Benedetti:2016rwo} for studies on the relation between CDT and Ho\v{r}ava-Lifshitz gravity. Indeed one would expect that matrix/tensor models which encode causality should be rich enough to encode various continuum universality classes within their phase diagram. A comparison of the scaling spectrum between continuum and discrete settings can provide a check of this expectation.\\
Such an improvement in accuracy is expected to arise from an improvement in the truncation. The critical exponents at the fixed points we uncovered here are not incompatible with a truncation principle that follows canonical scaling: As in other matrix and tensor models, an increasing order in powers of $A$ and $B$ as well as number of traces of an interaction decreases the scaling dimension of the associated coupling. For fixed-point candidates for which the deviation of the critical exponent from the canonical scaling dimensions is $\mathcal{O}(1)$, as it is in our case, robust truncations can be constructed by neglecting those interactions which are canonically highly irrelevant ones. \\
Further, setting up a coarse graining in matrix size $N$ leads to a violation of the $\text{U}(N')$ symmetry associated with each of the matrices (where $N'$ is a UV cutoff). The resulting Ward-identities have first been solved for matrix models in~\cite{Eichhorn:2014xaa}, where it was used that they become trivial in the tadpole-approximation. The latter is well-suited to characterize the fixed point in matrix models for 2d quantum gravity and might therefore also be viable in the present case of the $ABAB$ matrix model. Beyond the tadpole approximation, the solution of the Ward-identities has been explored in~\cite{Lahoche:2018ggd,Lahoche:2018hou,Lahoche:2019vzy}.\\

Beyond the characterization of the universality class in terms of the critical exponents, an understanding of the emergent geometries is desirable. Within the FRG, the composite-operator formalism~\cite{Pawlowski:2005xe,Igarashi:2009tj,Pagani:2016dof,Becker:2018quq,Houthoff:2020zqy} appears well-suited to provide access to, e.g., the scaling of geometric quantities. At the purely formal level, the evaluation of the corresponding flow equations for matrix/tensor models is not expected to be more challenging than for the flow equation for the effective action itself. The main challenge therefore lies in the identification of suitable operators, where  one would expect geometric information to be encoded in higher-order operators in the tensor/matrix formalism.

The relation of the $ABAB$ matrix model to a causal model of dynamical triangulations is an example to highlight a promising route to impose causality in matrix/tensor models: To take into account the presence of both timelike and spacelike edges in a triangulation, a multi-field approach seems indicated, see for instance~\cite{Kawabe:2020oiw}. The present article as well as the recent application of the FRG to a model of non-commutative spacetime with two fields~\cite{perezsanchez2020multimatrix} highlights that the FRG is well-suited to study such models. The present article lays the ground for extensions to higher-rank models.\\

Beyond the encoding of causality, multi-field models could also account for the interplay of quantum gravity with matter~\cite{Brezin:1989db,Bonzom:2012qx,Lahoche:2019orv}, providing yet another motivation to extend the present study to tensor models in the future.

\acknowledgments

The authors are grateful to an anonymous referee and R. Percacci for their remarks which led to an improvement of the manuscript. We also thank D.~Benedetti and J.~Th\"urigen for discussions. This work is supported by a research grant (29405) from VILLUM FONDEN.
A.~G.~A.~P.~is supported by the PRIME programme of the German Academic Exchange Service (DAAD) with funds from the German Federal Ministry of Education and Research (BMBF) and is thankful to CP3-Origins, University of Southern Denmark, for hospitality. ADP acknowledges CNPq under the grant PQ-2 (309781/2019-1) and FAPERJ under the ``Jovem Cientista do Nosso Estado'' program (E-26/202.800/2019) for support.

\bibliographystyle{JHEP}
\bibliography{refs}

\end{document}